\documentclass{ws-mpla}
\usepackage[super]{cite}
\usepackage{graphicx}
\usepackage{mathtools}
\usepackage{hyperref}
\usepackage{float}
\usepackage{orcidlink}
\begin{document}

\markboth{Sanjay Mandal, P.K. Sahoo}
{On The Temporal Evolution of Particle Production in $f(T)$ Gravity}

\catchline{}{}{}{} 

\title{On The Temporal Evolution of Particle Production in $f(T)$ Gravity}

\author{\footnotesize Sanjay Mandal\orcidlink{0000-0003-2570-2335}}

\address{Department of Mathematics,\\ Birla Institute of
Technology and Science-Pilani, \\ Hyderabad Campus, Hyderabad-500078,
India\\
sanjaymandal960@gmail.com}

\author{P.K. Sahoo\orcidlink{0000-0003-2130-8832}}

\address{Department of Mathematics,\\ Birla Institute of
Technology and Science-Pilani, \\ Hyderabad Campus, Hyderabad-500078,
India\\
pksahoo@hyderabad.bits-pilani.ac.in}

\maketitle

\pub{Received (22 July 2020)}{Accepted (05 October 2020)}

\begin{abstract}
The thermodynamical study of the universe allow particle production in modified $f(T)$
($T$ is the torsion scalar) theory of gravity within a flat FLRW framework for line element. The torsion scalar $T$ plays the same role as the Ricci scalar $R$ in the modified theories of gravity. We derived the $f(T)$ gravity models by taking $f(T)$ as the sum of $T$ and an arbitrary function of $T$ with three different arbitrary function. We observe that the particle production describes the accelerated expansion of the universe without a cosmological constant or any unknown ``quintessence" component. Also, we discussed the supplementary pressure, particle number density and particle production rate for three cases.

\end{abstract}

\ccode{PACS Nos.: 04.50.kd}

\section{Introduction}\label{I}
At the beginning of the 19th century, the General Theory of Relativity brought the revolution to the modern cosmology proposed by Albert Einstein. The Riemannian space-time formulates this theory based on the Levi-Civita connection, a torsion-free, and metric compatibility connection. It also helps us to understand the geodesic structure of the Universe. Later on, it faced problems like fine-tuning, cosmic co-incidence, initial singularity, cosmological constant, and flatness \cite{sahni/2000}, since modern cosmology is growing by a prominent number of accurate observations. Besides, the cosmological observation, such as type Ia supernovae \cite{Perlmutter/1999,Riess/1998}, cosmic microwave background (CMB) radiation \cite{Spergel/2003,Komatsu/2011}, large scale structure \cite{Tegmark/2004,Seljak/2005}, baryon acoustic oscillations \cite{Eisenstein/2005}, and weak lensing \cite{Jain/2003} confirms that currently, our Universe is going through an accelerated expansion phase that happens because of the highly negative pressure produced by the unknown form of matter and energy, called dark energy and dark matter. To overcome the above issues, researchers started to modify the Einstein's theory of relativity, and they ended up with several modified theories of gravity such as $f(R)$ gravity \cite{nojiri/2006}, $f(R,T)$ gravity \cite{harko/2011}, $f(T)$ gravity \cite{cai/2016}, $f(Q)$ gravity \cite{jimenez/2018}, $f(Q,T)$ gravity \cite{xu/2019}, etc. As a result, cosmologists found many interesting results such as Yousaf et al., studied the self-gravitating structures \cite{yousaf/2020}, gavastars \cite{yousaf/2020a}, Sahoo et al. \cite{sahoo/2020} studied the wormhole geometry, bouncing cosmology, and  accelerated expansion of the universe  using modified theories of gravity. The main advantage of the modified theories of gravity is that it successfully describes the late-time cosmic acceleration and the early time inflation. In the early stage of the Universe, there is a possibility of particle creation. In this study, we are focusing on particle production in the teleparallel gravity.
\newline
The $f(T)$ theories of gravity are the generalization of the Teleparallel Equivalent of General Relativity (TEGR), where $T$ is the torsion scalar \cite{Cai/2016}. TEGR was first presented by Einstein. In the context of $f(T)$ theories, the Reimann-Cartan space-time requires the torsional curvature to vanish. Furthermore, this type of space-time is constructed by the Weitzenb\"ock connection \cite{weit/1923,vinckers/2020}. TEGR is equivalent to GR; the reason is that both cases' action is the same except the surface term in TEGR. But, the physical interpretation is different from each other. The construction of gravitational Lagrangian in TEGR formulation was done in \cite{maluf/1994,andrade/1997}. TEGR is Lorentz invariant theories, whereas the modified $f(T)$ theories are not Lorentz invariant. Also, the motion equations in $f(T)$ gravity are not necessarily Lorentz invariant, because the certainty is that this theory's property explains the recent interests \cite{tamanini/2012,meng/2011,ferraro/2011,cruz/2014}. Moreover, the modification of TEGR was motivated by the $f(R)$ gravity theory. In the teleparallel gravity theory, we use the Weitzenb\"ock connection instead of the Levi-Civita connection, which uses $f(R)$ gravity to vanish the non-zero torsion curvature. Here, we would like to mention that the $f(T)$ gravity does not need the Equivalence principle because the Weitzenb\"ock connection describes its gravitational interaction. It is a simple modified theory compared to other modified theories because the torsion scalar $T$ contains only the first-order derivatives of the vierbeins. In contrast, Ricci scalar $R$ contains the second-order derivatives of the metric tensors. Recently, Mandal et al., studied the acceleration expansion of the Universe using the parametrization technique with presuming exponential and logarithmic form of $f(T)$ \cite{mandal/2020} in $f(T)$ gravity. They also studied a complete cosmological scenario of the Universe in $f(T)$ gravity, where they discussed the difference between General Relativity and Teleparallel gravity \cite{mandal/2020a} in $f(T)$ gravity. M. Sharif and S. Rani studied the dynamical instability ranges in Newtonian as well as post-Newtonian regimes considering power-law $f(T)$ model with anisotropic fluid in $f(T)$ gravity \cite{sharif/2014}.  Cai et al., studied the matter bounce cosmology using perturbation technique and they found a scale-invariant power spectrum, which is consistent with cosmological observations in $f(T)$ gravity \cite{cai/2011}. In \cite{sharif/2013}, the wormhole solutions with non-commutative geometry have been studied assuming power-law $f(T)$ model and a particular shape function in teleparallel gravity. Inflationary universe studied using power-law $f(T)$ function and logamediate scale factor in \cite{kazem/2017}, and constant-roll inflation studied in \cite{awad/2017}.
\newline
In the early stage of the universe, the possibility of particle creation has been discussed for curved space-time by Schrodinger \cite{schrodinger/1939}, Dewitt \cite{dewitt/1953}, Imamura \cite{imamura/1960}. Later, the first ever particle creation was treated by an external gravitational field by Parkar \cite{parkar/1968, parkar/1969}. In flat space-time, the unique vacuum state is identifying by the guidance of Lorentz invariance. Moreover, we do not have Lorentz symmetry in curved space-time. In general, there are more than one vacuum state exists in a curved space-time. Therefore, the particle creation idea becomes open to discuss, but it's physical interpretation becomes more difficult \cite{birrell/1982,ford/1997}. The interaction between the dynamical external gradients causes the particle creation from the vacuum. The particle creation produces negative pressure, so it is considered to explain the accelerated expansion of the universe and got some unexpected outcomes. Also, it might play the role of unknown gradients of the universe. In \cite{zimdahl/2001,qiang/2007} studied the particle creation with SNe Ia data. Singh \cite{csingh/2012}, and Singh and Beesham \cite{singh/2011,singh/2012} studied the particle creation with some kinematical tests in FLRW cosmology. The continuous creation of particle predicts the assumptions of standard Big Bang cosmology.\\
The thermodynamical study of black hole gives the fundamental relation between thermodynamics and gravitation \cite{bardeen/1973,bekenstein/1973,hawking/1975,gibbons/1977,bamba/2011c}. In GR, the relation between the entropy and the horizon area with the Einstein equation derives from the Clausius relation in thermodynamics \cite{jacobson/1995,elizalde/2008}. This idea is also used for other theories, mainly, the generalized thermodynamics laws and modified theories of gravity which are derived from the GR \cite{miaoa/2011,karami/2012}. Among the modified theories of gravity, $f(R)$ gravity got more attention on this framework. Thereby, one can obtained the gravitational field equation through the non-equilibrium feature of thermodynamics by using the Clausius approach. There are some work have been done in the thermodynamics of particle creation in $f(T)$ gravity theory \cite{setare/2013,bamba/2012,csingh/2014,salako/2013,bamba/2011c}.\\ 
In this work, we study the theoretical significance of particle creation in $f(T)$ gravity theory considering a flat FRW model. Assuming $f(T)$ as the sum of torsion scalar $T$ and an arbitrary function of torsion scalar $T$, we studied the thermodynamics of particle creation with $f(T)=0$ is a simple teleparallel gravity, $f(T)=A(-T)^q$ as power law gravity and $f(T)=A(1-e^{-qT})$ as exponential gravity. After that we discussed the behaviour of supplementary pressure $p_c$, particle number density $n$, and the particle creation rate $\psi$ for three models. Also we compared the effect of the cosmological pressure $p_m$ with the supplementary pressure $p_c$ for different values of equation of state parameter $\omega$ on particle creation.\\
This work is organised as follows. In Sec. \ref{II}, we discussed the thermodynamics of particle creation, which is followed by the overview of $f(T)$ gravity and it's field equations in Sec. \ref{III}. In Sec. \ref{IV}, we discussed three $f(T)$ gravity models. Finally, the results are summarized in Sec. \ref{VI}

\section{Thermodynamics of particle creation}\label{II}

If we assume the total number of particles in the universe to be conserved, the laws of thermodynamics can be expressed as 
\begin{equation}\label{3a}
d Q = d (\rho_{m}V)+p_{m}dV
\end{equation}
and 
\begin{equation}\label{3b}
T dS =p_{m}dV + d(\rho_{m}V)
\end{equation}
where $p_{m}$, $\rho_{m}$, $V$, $T$ and $S$ denote respectively the cosmological pressure, density, volume, temperature and entropy. Also, $dQ$ represent the heat exchange in the time interval $dt$. From (\ref{3a}) and (\ref{3b}), we further obtain,
\begin{equation}\label{3c}
d Q=TdS
\end{equation}
Eq. (\ref{3c}) reflects the fact that the entropy is a conserved quantity, since for  lose adiabatic system $dQ=0$. We now consider a scenario in which the total number of particles in the universe is not constant. Under this condition, Eq (\ref{3a}) gets modified to \cite{cpfrt}
\begin{equation}\label{3d}
d Q = d (\rho_{m}V)+p_{m}dV + (h/n) d(nV)
\end{equation}
where $N=nV$, $n$ being the number density of the particles and $h=(p_{m}+\rho_{m})$ the enthalpy per unit volume of the system. For an adiabatic system where $dQ=0$, (\ref{3d}) reads \cite{cpfrt}
\begin{equation}\label{3e}
d (\rho_{m}V)+p_{m}dV = (h/n) d(nV)
\end{equation}
In \cite{cpfrt}, the authors stated that in cosmology this change in the total number of particles in the universe can be understood as a transformation of gravitational field energy to the matter. \\
For an open thermodynamic system, Eq. (\ref{3e}) can be expressed as \cite{cpfrt}
\begin{equation}\label{3f}
d (\rho_{m}V) = -\left(p_{m}+p_{c} \right) dV
\end{equation}
where 
\begin{equation}\label{3g}
p_{c} = -(h/n)(dN/dV)
\end{equation}
represents supplementary pressure associated with the creation of particles \cite{cpfrt}. Note that negative $p_{c}$ indicate production of particles whereas positive $p_{c}$ implies particle annihilation and finally for $p_{c}=0$ the total number of particles is constant. Using Eq (\ref{3b}) and (\ref{3e}), it can also be shown that \cite{cpfrt}
\begin{equation}\label{3h}
S =S_{0}\left( \frac{N}{N_{0}}\right) 
\end{equation}
where $S_{0}$ and $N_{0}$ represent current values of these quantities. \\
Additionally, we assume the particles follow a barotropic equation of state and therefore can be written as 
\begin{equation}\label{3i}
p_{m}=(\omega)\rho_{m}
\end{equation}
where $-1\leq \omega \leq 1$ is the EoS parameter. The number density of particles is related to the density $\rho_{_{m}}$ as \cite{singh/2012}
\begin{equation}\label{3j}
n = n_{0} \left(\frac{\rho_{m}}{\rho_{0}} \right)^{\frac{1}{1+\omega}} 
\end{equation}
where $\rho_{0} \geq 0$ and $n_{0} \geq 0$ are the present values of density and particle number density respectively.\\
We now consider the matter creation rate to be defined as \cite{21}
\begin{equation}\label{3k}
\psi(t)=3 \beta n H
\end{equation}
where $0 \leq \beta \leq 1$ is assumed to be a constant and $\psi(t)$ represent the rate of particle creation and has a dimension of $t^{-1}$. $\psi$ can either be positive or negative depending on the creation or annihilation of particles. $\psi = 0$ indicate particle number being conserved in the universe. For cosmological matter following barotropic equation of state (Eq. \ref{3i}), the supplementary pressure $p_{c}$ can be expressed as \cite{cpfrt}
\begin{equation}\label{3l}
p_{c}=-\beta (\omega +1) \rho_{m}
\end{equation}

\section{Overview of $f(T)$ Gravity}\label{III}

Let us consider the extension of Einstein-Hilbert Lagrangian of $f(T)$ theory of gravity (which is similar to $f(R)$ gravity extension from the Ricci scalar $R$ to $R+f(R)$ in the action), namely the teleparallel gravity term $T$ to $T+f(T)$, where $f(T)$ is an arbitrary function of $T$ as
\begin{equation}
\label{a}
S=\frac{1}{16\pi G}\int[T+f(T)]e d^4x,
\end{equation}
where $e=det(e^i_\mu)=\sqrt{-g}$ and $G$ is the gravitational constant. Assume $k^2=8\pi G=M_p^{-1}$, where $M_p$ is the Planck mass.The gravitational field is defined by the torsion one as
\begin{equation}
\label{b}
T^{\gamma}_{\mu \nu}\equiv e^{\gamma}_i(\partial_{\mu} e^i_{\nu}-\partial_{\nu} e^i_{\mu}).
\end{equation}
The contracted form of the above torsion tensor is
\begin{equation}
\label{c}
T\equiv \frac{1}{4}T^{\gamma \mu \nu}T_{\gamma \mu \nu}+\frac{1}{2}T^{\gamma \mu \nu}T_{\nu \mu \gamma}-T^{\gamma}_{\gamma \mu}T^{\nu \mu}_{\nu}.
\end{equation}
By the variation of the total action $S+L_m$, here $L_m$ is the matter Lagrangian gives us the field equation for $f(T)$ gravity as
\begin{multline}
e^{-1}\partial_{\mu}(ee^{\gamma}_i S^{\mu \nu}_{\gamma})(1+f_T)-(1+f_T)e^{\lambda}_i T^{\gamma}_{\mu \lambda}S^{\nu \mu}_{\gamma}\\
+e^{\gamma}_i S^{\mu \nu}_{\gamma}\partial_{\mu}(T)f_{TT}+\frac{1}{4}e^{\nu}_i[T+f(T)]=\frac{k^2}{2} e^{\gamma}_iT^{(M)\nu}_{\gamma},
\label{d}
\end{multline}
where $f_T=df(T)/dT$, $ f_{TT}=d^2f(T)/dT^2$, the "superpotential `` tensor $S^{\mu \nu}_{\gamma}$ written in terms of cotorsion $K^{\mu \nu}_{\gamma}=-\frac{1}{2}(T^{\mu \nu}_{\gamma}-T^{\nu \mu}_{\alpha}-T^{\mu \nu}_{\alpha})$ as $S^{\mu \nu}_{\gamma}=\frac{1}{2}(K^{\mu \nu}_{\gamma}+\delta^{\mu}_{\gamma}T^{\alpha \nu}_{\alpha}-\delta^{\nu}_{\gamma}T^{\alpha \mu}_{\alpha})$ and $T^{(M)\nu}_{\gamma}$ represents the energy-momentum tensor to the matter Lagrangian $L_m$.
Now we consider a flat FLRW universe with the metric as
\begin{equation}
\label{e}
ds^2=dt^2-a^2(t)dx^{\mu} dx^{\nu},
\end{equation}
where $a(t)$ is the scale factor, which gives us
\begin{equation}
\label{f}
e^{i}_{\mu}=diag(1,a,a,a).
\end{equation}
Using equation (\ref{f}) into the field equation (\ref{d}), we get the modified field equation as follows
\begin{equation}\label{4a}
H^{2}=\frac{8 \pi G}{3}\rho_{m} -\frac{f}{6} + \frac{T f_{T}}{3},
\end{equation}
\begin{equation}\label{4b}
\dot{H}=-\left[\frac{4 \pi G (\rho_{m}+p_{m}+p_{c})}{1+f_{T}+2 T f_{T T}} \right] ,
\end{equation}
where $H\equiv \dot{a}/a$ be the Hubble parameter and "dot`` represents the derivative with respect to $t$. Here, $\rho_m$ and $p_m$ be the energy density and pressure of the matter content, $p_c$ be the supplementary pressure. Also, we have used
\begin{equation}
\label{g}
T=-6H^2,
\end{equation}
which holds for a FLRW Universe according to equation (\ref{c}).

\section{$f(T)$ gravity models}\label{IV}

In this section we shall investigate the temporal evolution of particle production in radiation ($\omega=1/3$) and dust universe ($\omega=0$) for various $f(T)$ gravity models with model parameters constrained from cosmological observations related to gravitational baryogenesis.\\
For the purpose of analysis, we shall assume a power law evolution of scale factor of the form 
\begin{equation}\label{5a}
a(t) = a_{0} t^{\left[ \frac{2}{3(1+\omega)}\right] }
\end{equation}
where $a_{0}>0$ is a constant.

\subsection{Simple Teleparallel gravity}

In simple teleparallel equivalent of general relativity \cite{12}, where $f(T)=0$ , for a universe composed of perfect fluid, the field equations (\ref{4a}) and (\ref{4b}) becomes
\begin{equation}\label{6a}
H^{2}=\frac{8 \pi G}{3}\rho_{m}
\end{equation}
\begin{equation}\label{6b}
\dot{H}=-4 \pi G (\rho_{m}+p_{m}+p_{c}) 
\end{equation}
Substituting (\ref{5a}) in (\ref{6a}), we obtain the expression of density $\rho_{m}$ as 
\begin{equation}\label{6c}
\rho_{m}=\frac{4}{3}\left( \frac{1}{t^{2}(1+\omega)^{2}}\right) 
\end{equation}
The expression of supplementary pressure $p_{c}$, particle number density $n$ and particle creation rate $\psi$ are obtained respectively as
\begin{equation}\label{6d}
p_{c}=\frac{4}{3} \left( \frac{\beta (1+\omega)}{t^{2}(1+\omega)^{2}}\right) 
\end{equation}
\begin{equation}\label{6e}
n=\left[ \frac{4}{3}\left( \frac{1}{t^{2}(1+\omega)^{2}}\right)\right] ^{\frac{1}{(1+\omega)}}
\end{equation}
\begin{equation}
\psi = 3 \beta \left[ \frac{2}{3 t(1+\omega)}\right] \times  \left[ \frac{4}{3}\left( \frac{1}{t^{2}(1+\omega)^{2}}\right)\right] ^{\frac{1}{(1+\omega)}}
\end{equation}
\begin{figure}[H]
  \centering
  \includegraphics[width=11 cm]{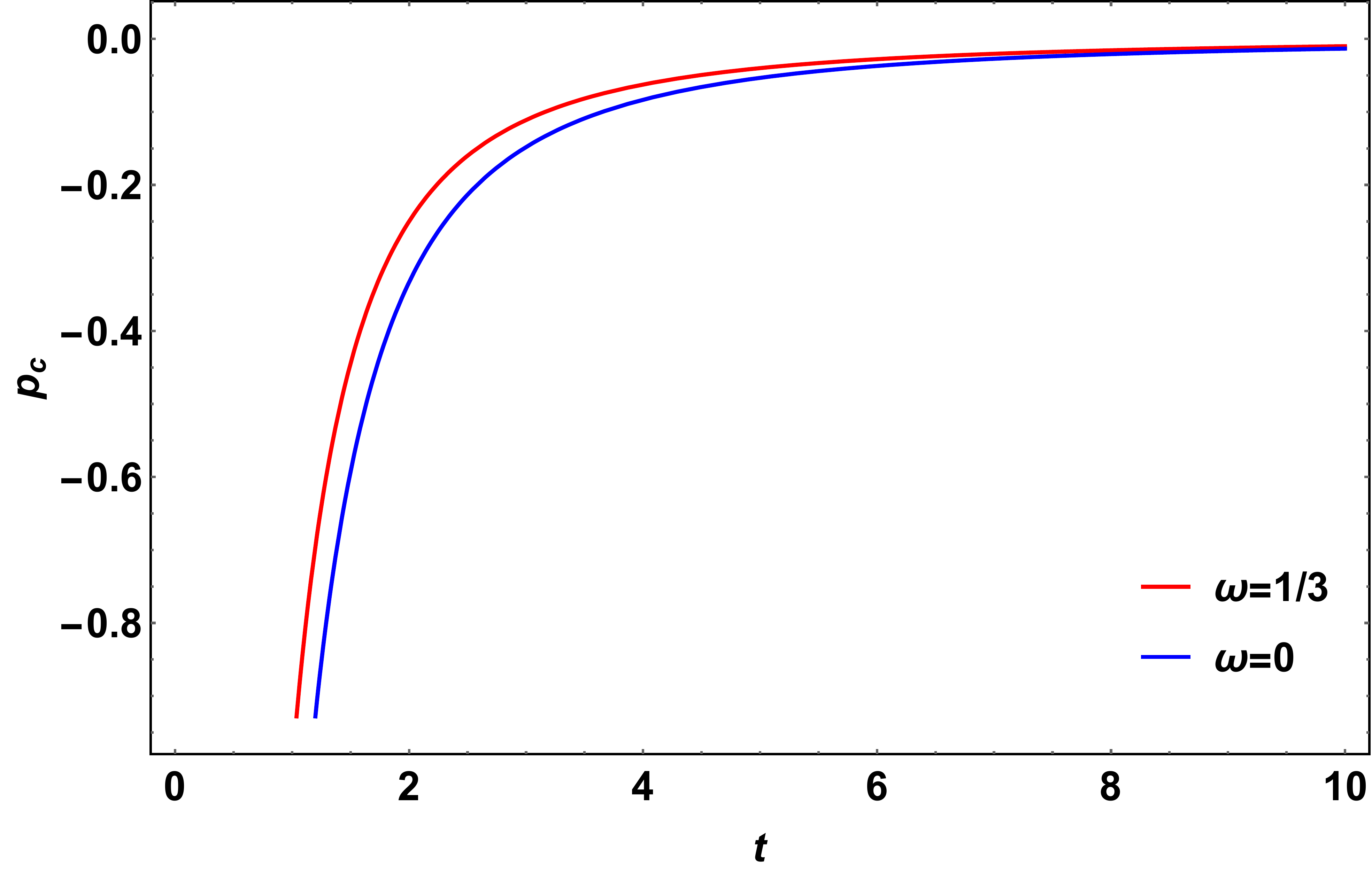}
  \caption{The behaviour of supplementary pressure $p_c$ with respect to cosmic time $t$ for $\omega=\frac{1}{3}, \omega=0$ and $\beta=1$.}
  \label{f1}
\end{figure}
\begin{figure}[H]
  \centering
  \includegraphics[width=11 cm]{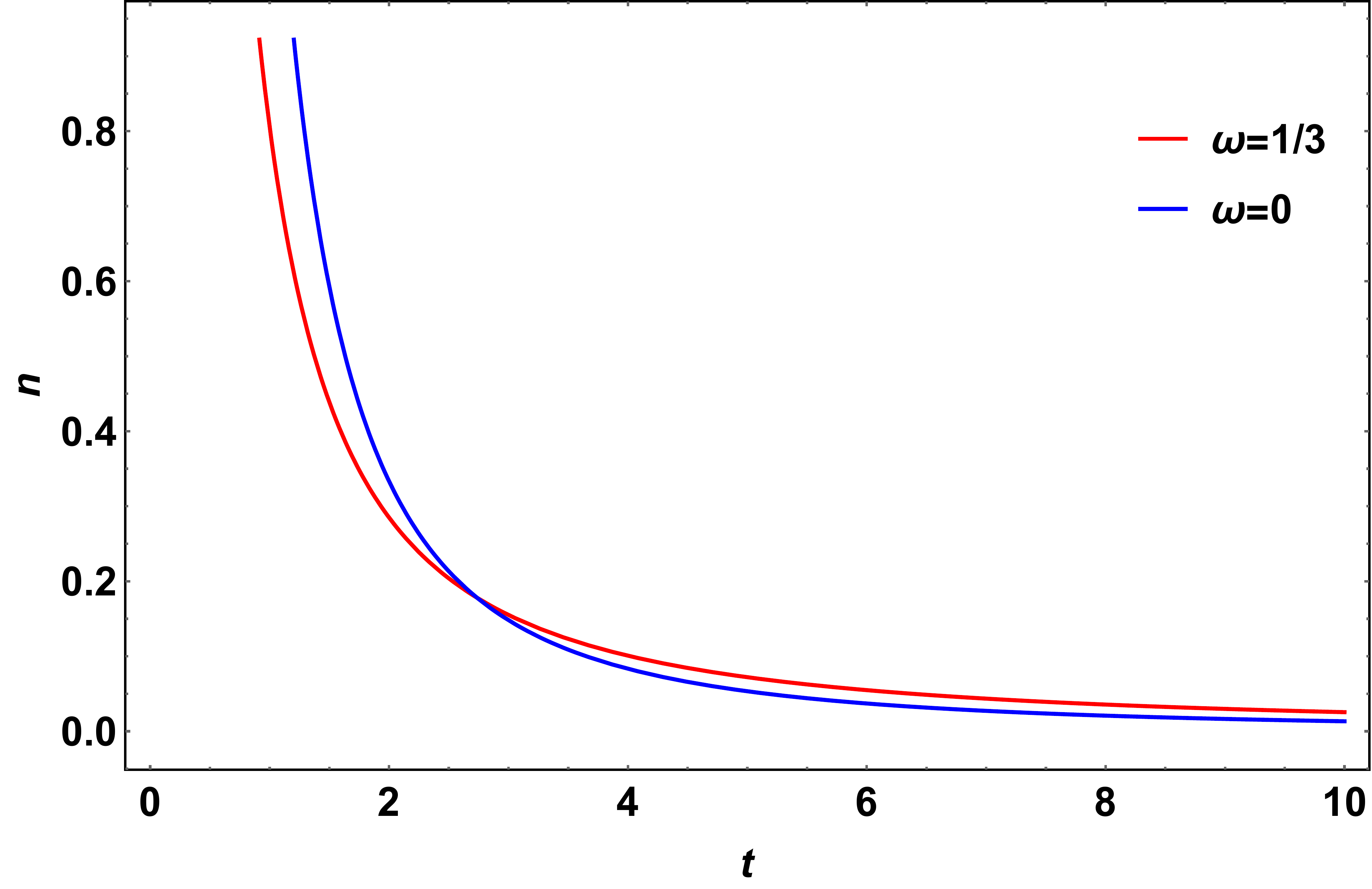}
  \caption{The behaviour of particle number density $n$ with respect to cosmic time $t$ for $\omega=\frac{1}{3}, \omega=0$ and $\beta=1$.}
  \label{f2}
\end{figure}
\begin{figure}[H]
  \centering
  \includegraphics[width=11 cm]{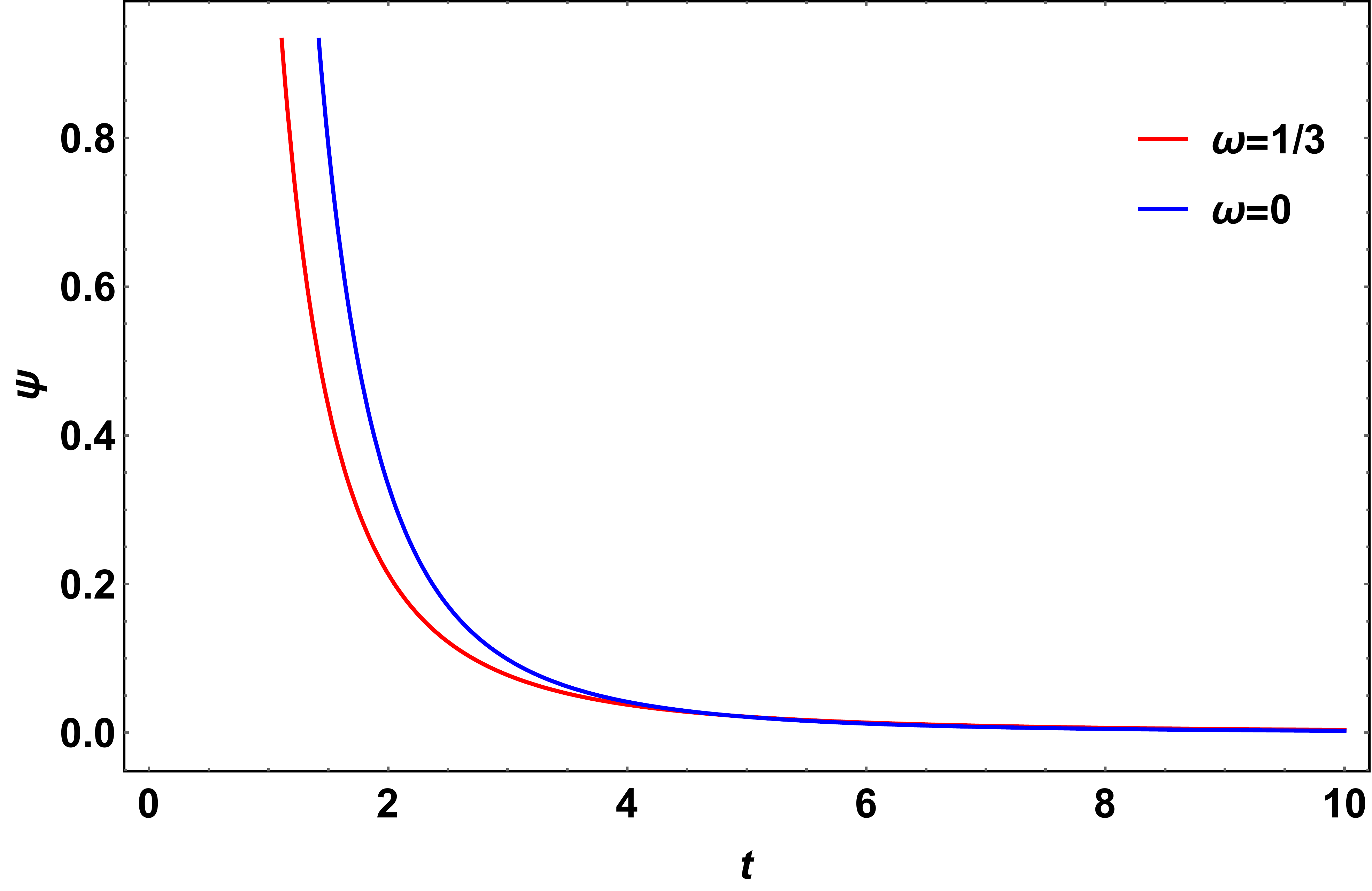}
  \caption{The behaviour of particle creation rate $\psi$ with respect to cosmic time $t$ for $\omega=\frac{1}{3}, \omega=0$ and $\beta=1$.}
  \label{f3}
  \end{figure}
  
\subsection{Power Law Gravity}

The power law model of Bengochea and Ferraro \cite{16} reads 
\begin{equation}\label{7a}
f(T)=  A(-T)^{q}
\end{equation}
where $A$ is a constant and $q>1$. In \cite{oiko}, the authors reported viable baryon-to-entropy ratio for $A=-10^{-7} \texttt{or} -10^{-6}$ and $q\gtrsim 4.8$. However, other values of the model parameters could also yield viable estimates of baryon-to-entropy ratio. Nonetheless, we restrict ourselves to the values $A=-10^{-7}$ and $q=5$ for the present analysis. Substituting (\ref{5a}) and (\ref{7a}) in (\ref{4a}) and (\ref{4b}), the expression of density $\rho_{m}$ reads 
\begin{equation}\label{7b}
\rho_{m}=A6^{(q-1)}(1-18q)t^{-2q}\left[ \frac{2}{3(1+\omega)}\right]^{2q}
\end{equation} 
The expression of supplementary pressure $p_{c}$, particle number density $n$ and particle creation rate $\psi$ for the power law gravity are obtained respectively as
\begin{equation}\label{7c}
p_{c}=-\beta (1+\omega)A6^{(q-1)}(1-18q)t^{-2q}\left[ \frac{2}{3(1+\omega)}\right]^{2q} 
\end{equation}
\begin{equation}\label{7d}
n=\left[ A6^{(q-1)}(1-18q)t^{-2q}\left[ \frac{2}{3(1+\omega)}\right]^{2q}\right] ^{1/(1+\omega)}
\end{equation} 
\begin{equation}\label{7e}
\psi = 3 \beta \left[ \frac{2}{3 t(1+\omega)}\right] \times  \left[ A6^{(q-1)}(1-18q)t^{-2q}\left[ \frac{2}{3(1+\omega)}\right]^{2q}\right] ^{1/(1+\omega)}
\end{equation}

\begin{figure}[H]
  \centering
  \includegraphics[width=11 cm]{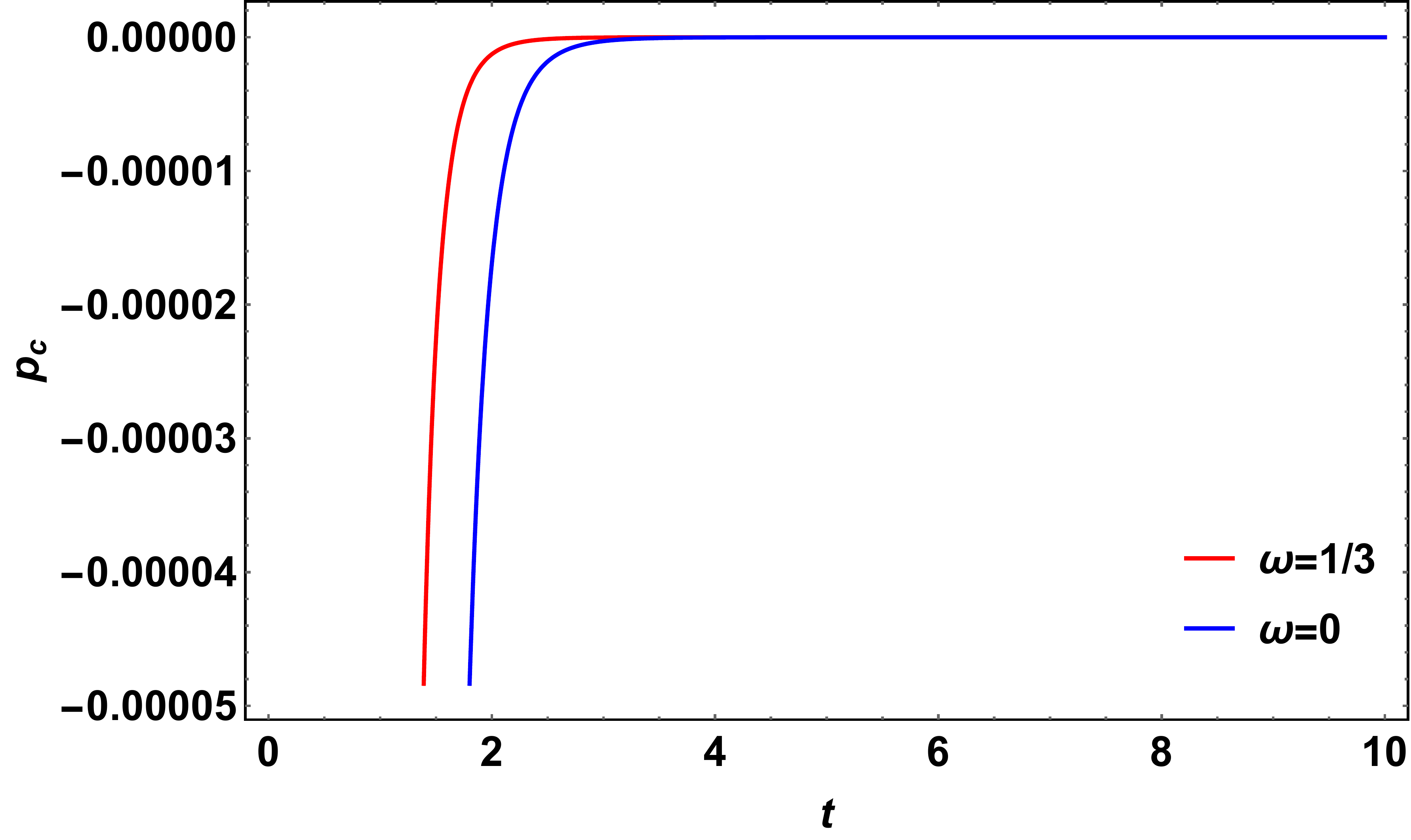}
  \caption{The behaviour of supplementary pressure $p_c$ with respect to cosmic time $t$ for $\omega=\frac{1}{3}, \omega=0$ and $\beta=1, q=5, A=-10^{-7}$}
  \label{f4}
\end{figure}
\begin{figure}[H]
  \centering
  \includegraphics[width=11 cm]{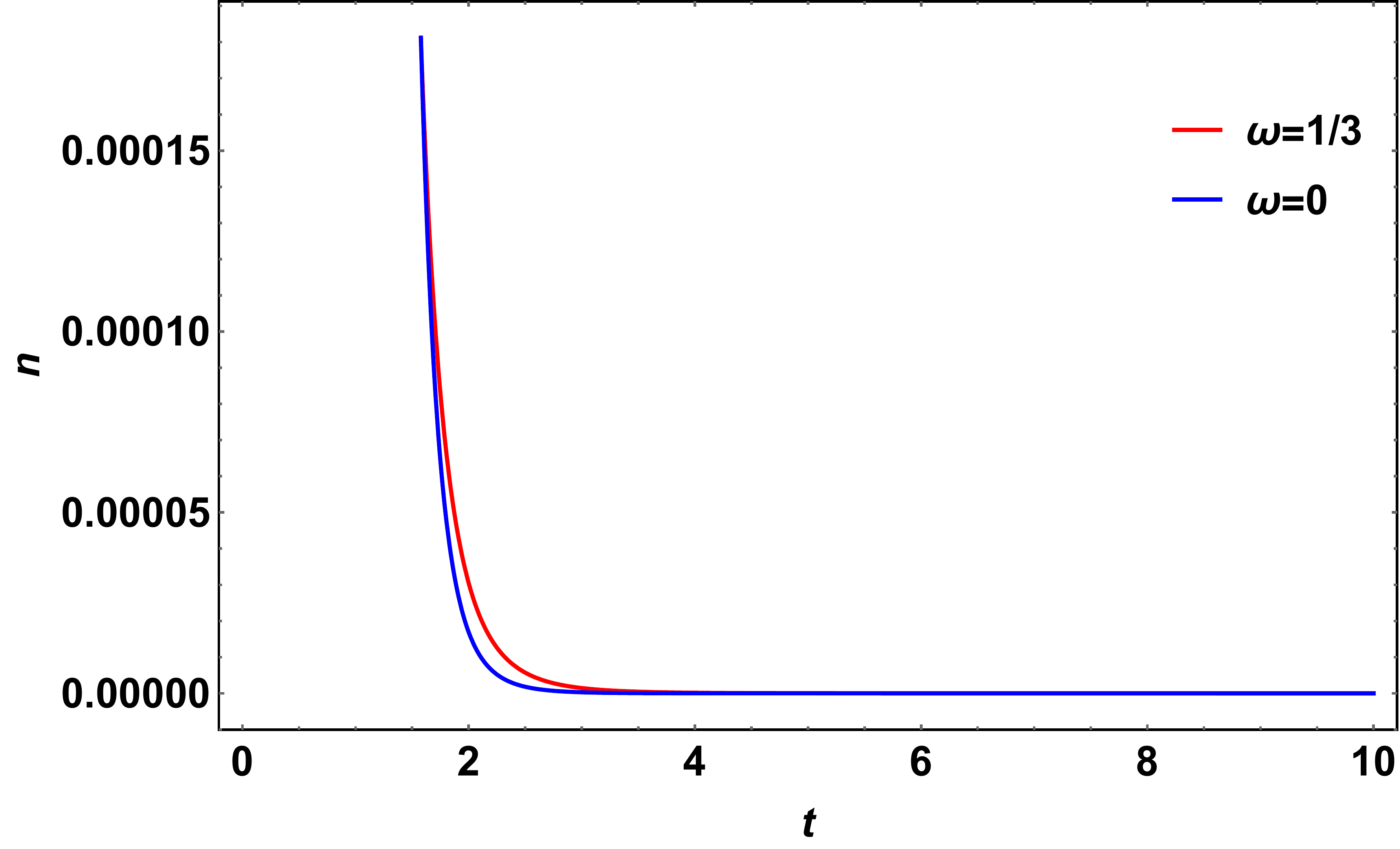}
  \caption{The behaviour of particle number density $p_c$ with respect to cosmic time $t$ for $\omega=\frac{1}{3}, \omega=0$ and $\beta=1, q=5, A=-10^{-7}$.}
  \label{f5}
\end{figure}
\begin{figure}[H]
  \centering
  \includegraphics[width=11 cm]{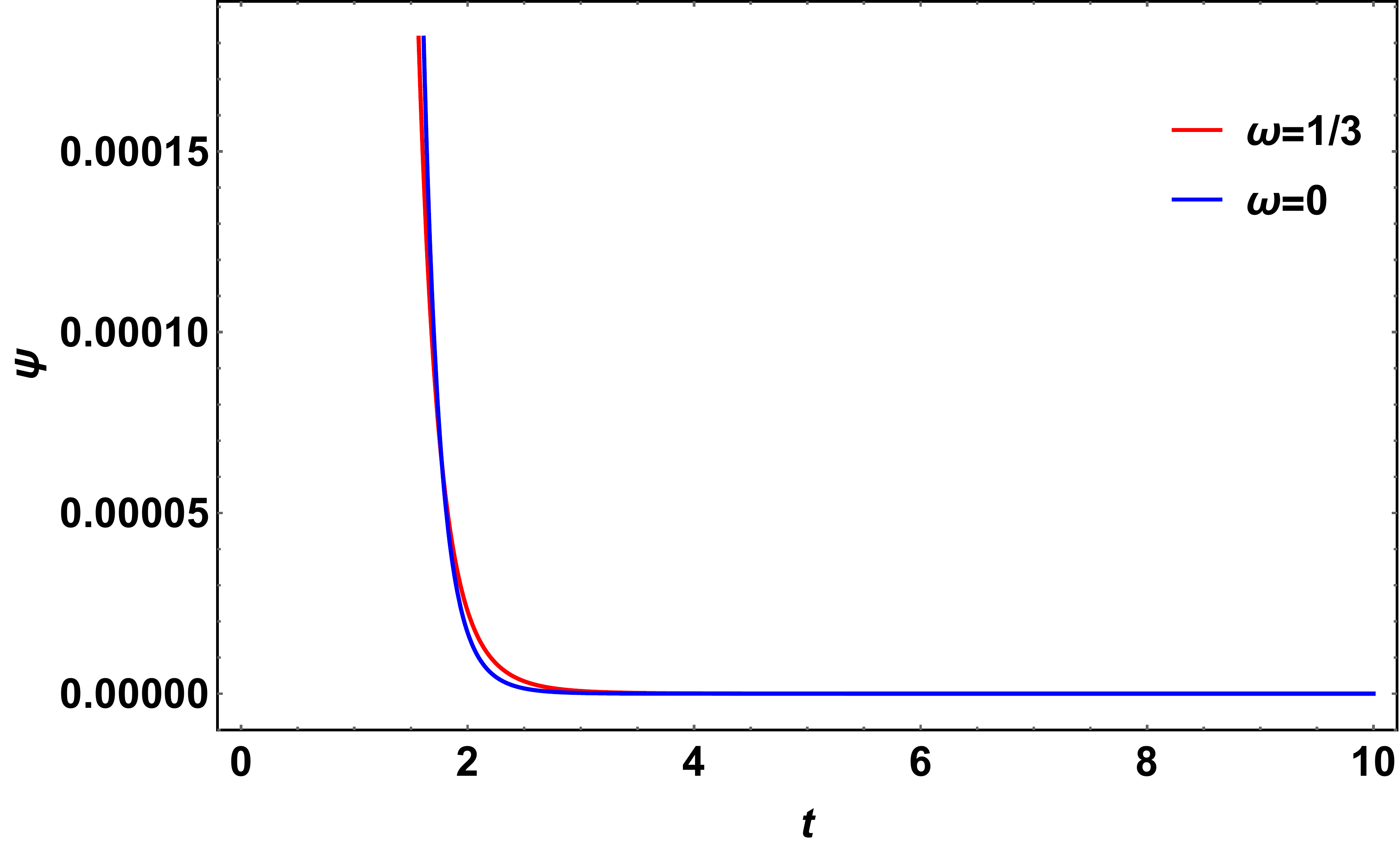}
  \caption{The behaviour of particle creation rate $\psi$ with respect to cosmic time $t$ for $\omega=\frac{1}{3}, \omega=0$ and $\beta=1, q=5, A=-10^{-7}$.}
  \label{f6}
\end{figure}

\subsection{Exponential Gravity}

The exponential $f(T)$ model given in \cite{23} reads 
\begin{equation}\label{8a}
f(T)=A (1-e^{-q T})
\end{equation}
where $A$ and $q$ are model parameters. In \cite{oiko} the authors reported a wide range of values of $A$ and $q$ for which a viable baryon-to-entropy ratio could be realized. However, we shall work with $A=1$ and $q=10^{-10}$ as these values were used in \cite{oiko}  to fit the baryon-to-entropy ratio with observations. Substituting (\ref{5a}) and (\ref{8a}) in (\ref{4a}) and (\ref{4b}), the expression of density $\rho_{m}$ reads 
\begin{equation}\label{8b}
\rho_{m} =\left[  \frac{18 A \left[ \frac{2}{3(1+\omega)}\right]^{2}q e^{ \frac{6\left[ \frac{2}{3(1+\omega)}\right]^{2}q}{t^{2}}}}{t^{2}}\right] 
\end{equation}
The expression of supplementary pressure $p_{c}$, particle number density $n$ and particle creation rate $\psi$ for the exponential gravity are obtained respectively as
\begin{equation}\label{8c}
p_{c}=-\beta (1+\omega) \left[ \frac{18 A \left[ \frac{2}{3(1+\omega)}\right]^{2}q e^{ \frac{6\left[ \frac{2}{3(1+\omega)}\right]^{2}q}{t^{2}}}}{t^{2}}\right] 
\end{equation}
\begin{equation}\label{8d}
n=\left[  \frac{18 A \left[ \frac{2}{3(1+\omega)}\right]^{2}q e^{ \frac{6\left[ \frac{2}{3(1+\omega)}\right]^{2}q}{t^{2}}}}{t^{2}}\right]^{1/(1+\omega)}
\end{equation}
\begin{equation}\label{8e}
\psi = 3 \beta \left[ \frac{2}{3 t(1+\omega)}\right] \times \left[\left(  \frac{18 A \left[ \frac{2}{3(1+\omega)}\right]^{2}q e^{ \frac{6\left[ \frac{2}{3(1+\omega)}\right]^{2}q}{t^{2}}}}{t^{2}}\right)^{1/(1+\omega)} \right] 
\end{equation}

\begin{figure}[H]
  \centering
  \includegraphics[width=11 cm]{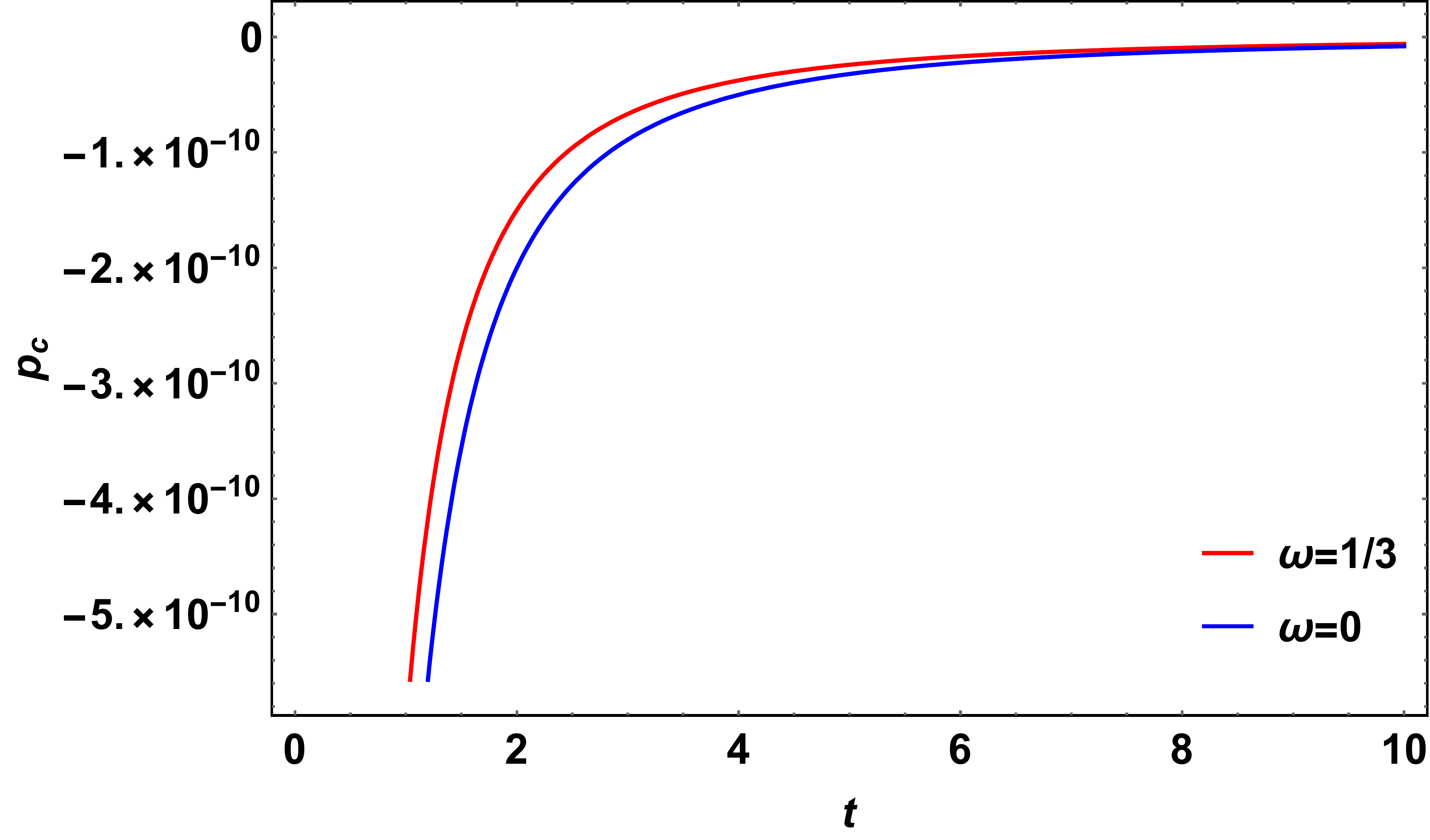}
  \caption{The behaviour of supplementary pressure $p_c$ with respect to cosmic time $t$ for $\omega=\frac{1}{3}, \omega=0$ and $\beta=1, q=10^{-10}, A=1$.}
  \label{f7}
\end{figure}
\begin{figure}[H]
  \centering
  \includegraphics[width=11 cm]{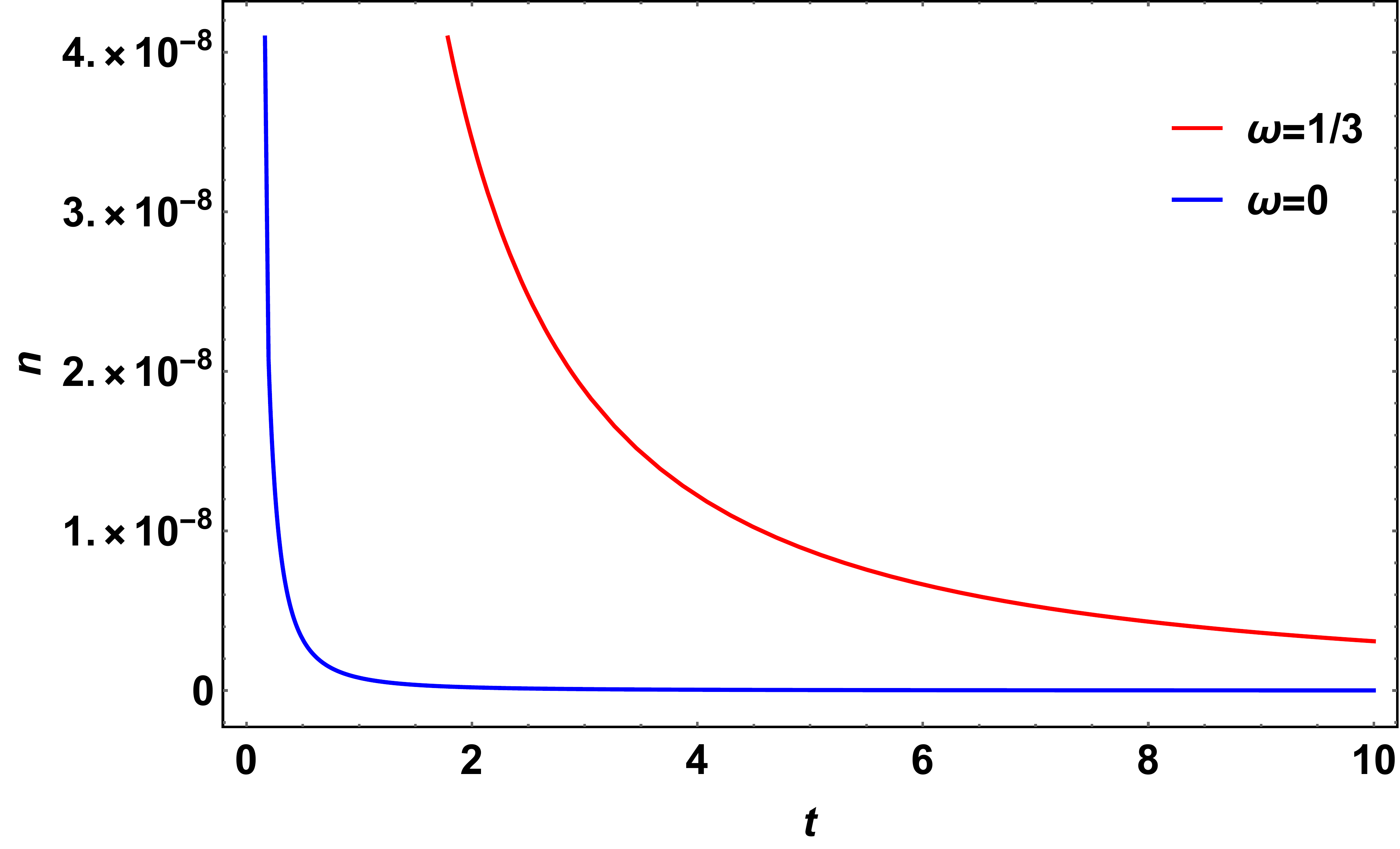}
  \caption{The behaviour of particle number density $n$ with respect to cosmic time $t$ for $\omega=\frac{1}{3}, \omega=0$ and $\beta=1, q=10^{-10}, A=1$.}
  \label{f8}
\end{figure}
\begin{figure}[H]
  \centering
  \includegraphics[width=11 cm]{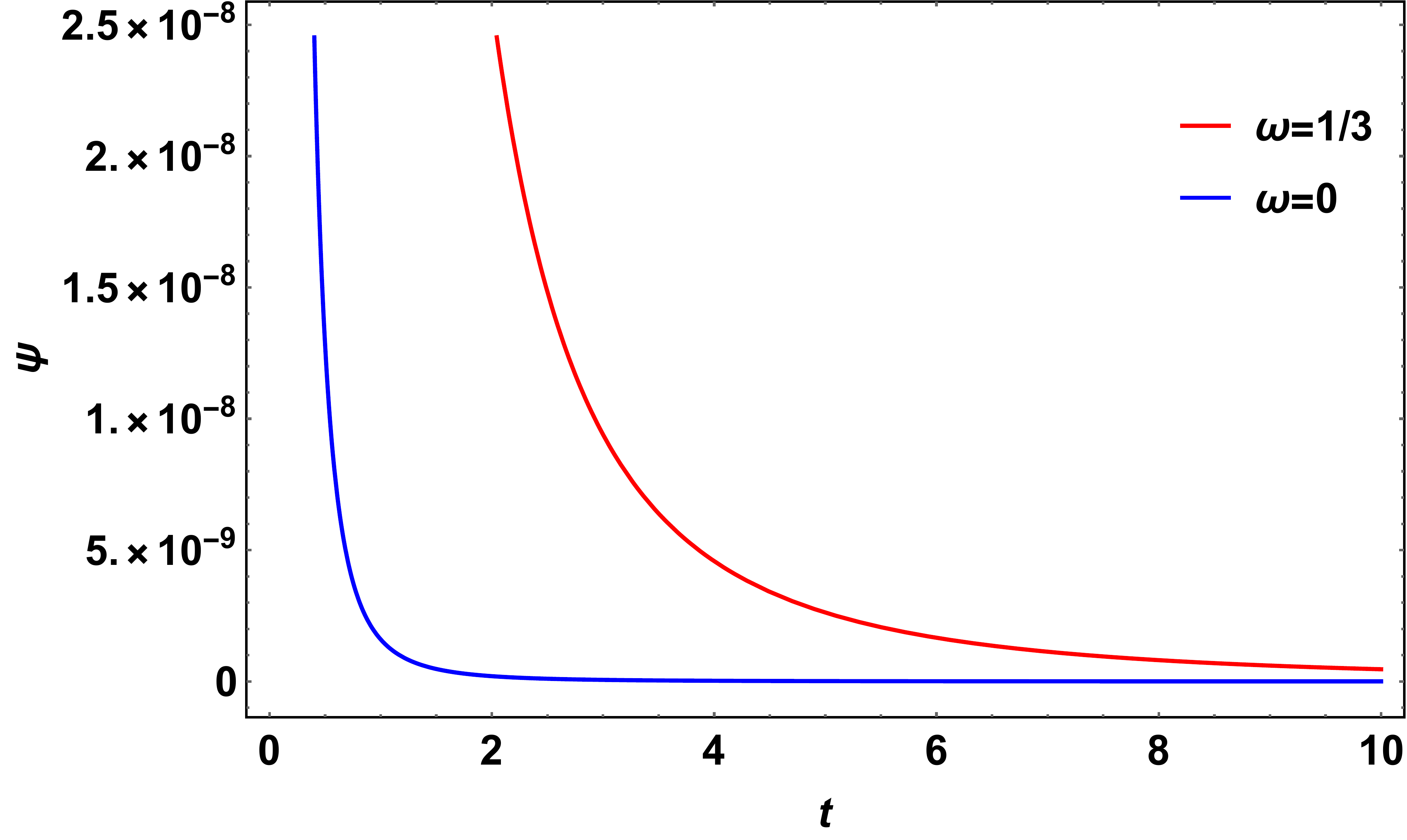}
  \caption{The behaviour of particle creation rate $\psi$ with respect to cosmic time $t$ for $\omega=\frac{1}{3}, \omega=0$ and $\beta=1, q=10^{-10}, A=1$.}
  \label{f9}
\end{figure}

\section{Discussion of Outcomes and Conclusions}\label{VI}

In this article, we have studied the thermodynamics of an open system with particle creation of a flat FLRW universe in $f(T)$ theory of gravity. We have constructed three cosmological models by assuming suitable functions for $f(T)$ as $f(T)=0, f(T)=A(-T)^q , f(T)=A(1-e^{-qT})$ and the particle creation  rate $\psi$. To analyze our models, we have considered the power law evolution of the scale factor and studied the behaviour of physical quantities (i.e. the supplementary pressure $p_c$, particle number density $n$, and the particle creation rate $\psi$) through their graphical representations with respect to cosmic time $t$ and some fixed values of $\beta$ in various phase of the evolution of the universe. And, details of our cosmological models discussed in the following.

In our simple Teleparallel gravity model, we have considered the minimal coupling between matter and geometry. In Fig. \ref{f3}, profiles of $\psi$ have been shown. From Fig. \ref{f3}, one can easily observe that the rate of particle creation is high in the early time and tends to zero when $t$ tends to infinity. But, the number of particle in the universe increases with cosmic time $t$ shown in Fig. \ref{f2}. The supplementary pressure $p_c$ has higher negative which shows that the particle production is high during the early stage and tends to zero when $t$ tends infinity, in Fig. \ref{f1}. From this model we have concluded that the evolution of the universe depends on the contribution of the particle production.

In power law gravity and exponential gravity models, we have considered the non-minimal coupling between matters. The profiles of $\psi, n$ and $p_c$ have been shown for the corresponding models. In Fig. \ref{f6},\ref{f9}, the particle creation rate $\psi$ is high in the early stage and it tends to zero as cosmic time $t$ tends to infinity. Also, the particle number density n in Fig. \ref{f5},\ref{f8} goes to zero as cosmic time $t$ goes to infinity which concluded that the expansion rate overcomes the rate particle creation as the supplementary pressure in Fig. \ref{f4},\ref{f7} is negative throughout the evolution of the universe in different phases. The density parameters in three models shows that the universe is open in the presence of particle creation in $f(T)$ theory of gravity.

In summary, we have studied the cosmological models with particle production in $f(T)$ theory of gravity to explore the current accelerated phenomenon of the universe. We have found that the particle creation produces negative pressure which may derive the accelerated expansion of the universe and play the role of unknown matter called ``dark energy" in $f(T)$ theory of gravity. We may expect that the particle creation process be a  constraint for the unexpected observational outcomes. The new fact about this article is that the particle creation is studied by the thermodynamics approach in $f(T)$ theory of gravity.

\textbf{Acknowledgements}\  S.M. acknowledges Department of Science \& Technology (DST), Govt. of India, New Delhi,
for awarding Junior Research Fellowship (File No.
DST/INSPIRE Fellowship/2018/IF180676). PKS acknowledges DST, New Delhi, India for providing facilities through DST-FIST lab, Department of Mathematics, BITS-Pilani, Hyderabad Campus where a part of this work was done. The authors  thank S. Bhattacharjee for stimulating discussions. We are very much grateful to the honorable referee and the editor for the illuminating suggestions that have significantly improved our work in terms of research quality and presentation.

\end{document}